\documentclass[12pt,preprint]{aastex}

\def\hide#1{}

\begin{document}
\title{ Dependence of kHz QPO Properties on 
Normal-Branch Oscillation Phase in Scorpius X-1}

\author{Wenfei Yu\altaffilmark{1,2}, Michiel van der Klis\altaffilmark{1} 
and Peter G. Jonker\altaffilmark{1}}

\altaffiltext{1}{Astronomical Institute, ``Anton Pannekoek'', 
	University of Amsterdam, and Center for High Energy Astrophysics, 
	Kruislaan 403, 1098 SJ Amsterdam, The Netherlands. E-mail:
yuwf@astro.uva.nl; michiel@astro.uva.nl; peterj@astro.uva.nl}

\altaffiltext{2}{Laboratory for Cosmic Ray and High Energy Astrophysics, 
	Institute of High Energy Physics, Beijing, 100039, China. 
	E-mail: yuwf@astrosv1.ihep.ac.cn}

\begin{abstract}
We analysed RXTE data of Sco X-1 which show kHz quasi-periodic
oscillations (QPOs) and the $\sim$6--8 Hz normal-branch oscillation
(NBO) simultaneously.  Using power spectra of 0.03--0.5~s data
segments, we find that both the upper kHz QPO frequency $\nu_2$ and
the ratio of lower to upper kHz QPO amplitude are anticorrelated to
variations in the X-ray count-rate taking place on the NBO time
scale. The frequency dependence is similar to (but probably weaker
than) that found on longer time scales, but the power ratio dependence
is {\it opposite} to it. A model where radiative stresses on the disk
material, modulated at the NBO frequency, lead to changes in 
$\nu_2$ can explain the
data; this implies some of the NBO flux changes originate from inside
the inner disk radius. We discuss how these findings affect our understanding of kHz
QPOs and of the low-frequency variability of low-mass X-ray binaries.
\end{abstract}


\section{Introduction}
Observations with the Rossi X-ray Timing Explorer (RXTE) of
neutron-star low-mass X-ray binaries (LMXBs) have led to the discovery
of kilohertz quasi-periodic oscillations (kHz QPOs) in about 20
sources (van der Klis 2000 for a review).  The kHz QPO frequency is
always correlated with position in the tracks traced out by these
sources in X-ray color-color diagrams. This usually implies kHz QPO
frequency is correlated with X-ray count rate, but anticorrelation
also occurs (Wijnands et al. 1997, Homan et al. 2001, this paper). On
time scales longer than hours to days the count-rate/frequency
correlations usually break down (e.g., M\'endez 1999; see also van der
Klis 2001). Usually a pair of kHz QPO peaks is observed (called lower
and upper kHz QPO in order of increasing frequency), with a frequency
separation $\Delta\nu$ that varies in anticorrelation with, but by
much less than, the kHz QPO frequencies themselves. In Sco X-1 (van
der Klis et al. 1996, 1997) for variations in frequency by
$\sim$500\,Hz, $\Delta\nu$ varies by $\sim$70\,Hz.

Numerous models have been proposed to explain the kHz QPOs (see van
der Klis 2000). Most associate one of the two kHz QPO frequencies with
the Keplerian frequency at the inner edge of the accretion
disk. Miller, Lamb \& Psaltis (1998) describe how the kHz QPO frequency 
can be affected by radiative stresses (both drag and radial force).
The $\sim$6\,Hz normal-branch oscillation (NBO) seen in
the most luminous LMXBs, the Z sources, is probably related to 
near-Eddington accretion (e.g. Hasinger et al. 1990;
but see Wijnands, van der Klis and Rijkhorst 1999). Several
models have been proposed for the NBO (see \S\ref{disc}).

In the Z source Sco X-1, the NBO occurs simultaneously with kHz QPOs
(van der Klis 1996, 1997).  At a fractional rms amplitude of 3-5\% it
dominates the 4-8~Hz X-ray variability. If the NBO-modulated flux
originates from within or near the inner disk radius then the
associated 6-Hz modulation of the drag on the inner disk could
modulate the orbital frequency there. If kHz QPO
frequencies are indeed related to orbital frequency, a modulation with
NBO phase would be expected. If, on the other hand, the 6-Hz
modulation originates further out, or the kHz QPO frequencies have
another origin, then such modulation of the kHz QPO frequencies would
not be predicted.

So, clues to the origin of both kHz QPOs and slower fluctuations can
be obtained by monitoring the kHz QPOs on subsecond time scales.  In
this Letter, we perform the first study of this kind. We investigate
the response of the kHz QPOs to the flux variations due to the NBO in
Sco X-1. We find, that while the response of the kHz QPO frequencies
to the 6-Hz flux variations is similar to (but probably less strong
than) that to slower variations in flux, the power ratio of the two
kHz peaks varies in the {\it opposite} sense. We discuss how these
findings can help our understanding of both NBO and kHz QPOs.

\section{Analysis and Results}
We analyzed 2--60 keV high time-resolution data taken of Sco X-1
between Feb 1996 and July 1998 with the RXTE Proportional Counter
Array (PCA). Based on 1/16~Hz resolution Fourier
power spectra of individual $\sim$2--4 ks continuous data segments, we selected
segments where the upper kHz QPO was significant at $>4\sigma$, and
the NBO was strong (peak Leahy power $>$5.0), narrow (FWHM $\sim$2--4
Hz), and between 6 and 8 Hz (Table 1).  We used data from all active
PCUs (3-5).  The 2--60 keV count rate was $\sim2.1\times{10}^{4}$ c/s
per PCU. Lower kHz QPOs were detected in about 1/3 of the segments.
 
We made FFTs with length $\delta{t}=1/32$ s and time resolution 1/4096
s. The frequency resolution of the resulting power spectra is 32 Hz
and the Nyquist frequency 2048 Hz. We compared the counts in
non-overlapping pairs of successive power spectra. Let the number of
counts in the $k$th pair of successive power spectra be $C^{k}_{H}$
for the higher count-rate power spectrum of the pair, and $C^{k}_{L}$
for the lower count-rate one. We then selected those pairs for which
the statistic $S_k\equiv(C^k_H-C^k_L)/\sqrt{(C^k_H+C^k_L)/2}$ exceeds
a given pre-set value.  $S_k$ is a measure for the significance of the
count-rate difference between the two power spectra in each pair. We
used those selected pairs to construct a high-rate sample of power
spectra, with corresponding selected $C^k_H$ values, and a low-rate
sample, with corresponding $C^k_L$ values. Two criteria were used in
the pair selection: $1.0\leqslant S_k<2.5$ and $S_k\geqslant2.5$. We
also combined the two to get the sample corresponding to
$S_k\geqslant1.0$.  We averaged the high-rate and low-rate power
spectra separately and fitted the average power spectra in the
300--1800 Hz range with two Lorentzians (for the two kHz QPOs) plus a
line (for the dead-time modified Poisson noise). All the uncertainties
were determined using $\Delta{\chi}^2=1.0$.

The difference $C^k_H-C^k_L$ is related to the Fourier power around
frequency $1/2\delta{t}$.  Hence, with $\delta{t}=1/32$\,s we sample
the fluctuations near 16 Hz.  To cover lower frequencies we
constructed longer segments of duration $N\delta{t}$ where $N = 1, 2,
3, ... 32$, which sample frequencies of $\sim16/N$~Hz. We calculated
the total counts and the average of the $N$ power spectra in each
$N\delta{t}$ wide segment, and constructed high-rate
and low-rate average power spectra for these longer segments in the
same way as above for $N=1$. The frequency and amplitude of the kHz
QPOs were then determined by fitting each high-rate and low-rate
average power spectrum for each N as described above.

Counting statistics of course contribute to the count differences
between the segments, but at near-NBO frequencies the NBO by far
dominates the variability.  Monte-Carlo simulations show that the
contribution of counting statistics to the count differences is
$\sim$2\% for $1\leqslant S_k < 2.5$, and $\sim$8\% for
$S_k\geqslant2.5$; around the NBO frequency these numbers drop to
$\sim$0.5\% and $\sim$6\%, respectively. With the exception of Figure
3, where we report absolute count rates, the average count differences
and derived quantities reported in this paper have all been corrected
for this counting-statistics bias associated with our selection
procedure.

The kHz QPOs in the average power spectra were significant at
$>3\sigma$, except the lower kHz QPO in the high-rate sample for N=1,
2, 3 and 4 ($\sim1.5\sigma$). Below we refer to the frequencies of the
lower and upper kHz QPO as $\nu_1$ and $\nu_2$and mark parameters for
the high-rate and low-rate samples with ``H'', and ``L''.  Fig.~1,
shows results for ${S}_{k}\geqslant1.0$. Fig.\,1a is the average power
spectrum of the observations, showing the NBO peak.  Fig.\,1b shows
the difference $\nu_{2L}-\nu_{2H}$ between $\nu_{2}$ in the low- and
high-rate samples. Near the NBO frequency, $\nu_{2L}$ is 10--20 Hz
larger than $\nu_{2H}$. As the counting-statistics-corrected average
count-rate difference between the high and the low-rate samples
$C_H-C_L$ is different at each frequency (i.e., each $N$), we show in
Fig.\,1c $\nu_{2L}-\nu_{2H}$ normalized to count-rate difference (in
units of counts per 1/32 s). Also when normalized this way, there is a
clear anticorrelation between $\nu_2$ and count rate.  The lower kHz
frequencies $\nu_{1H}$ and $\nu_{1L}$ are consistent with being the
same near the NBO frequency, and the frequency separation $\Delta\nu$
between the two peaks is larger in the low-rate sample:
$\Delta\nu_{L}>\Delta\nu_{H}$. However, as shown in Fig.\,1d, this is
only a 2-$\sigma$ effect and could be affected by the strength
variations in the peaks reported below.

Fig.\,2 shows the average high- and low-rate power spectra for N=2 and
${S}_{k}\geqslant2.5$. The raw measured integrated Leahy powers are
2.5$\pm$0.5 and 7.2$\pm$0.5 at low count rate and 0.09$\pm$0.39 and
4.7$\pm$0.4 at high count rate for lower and upper kHz peak,
respectively. The dashed line marks $\nu_{2}$ of the low-rate sample,
$1092.9\pm2.0$ Hz. At high count rate only the upper kHz QPO, at
$1070.5\pm1.5$ Hz, is significant. Taken at face value, Fig. 2
suggests that on the NBO time scale, going to higher count rate, the
lower kHz QPO weakens by a large factor, while the upper one remains
comparatively constant in strength (it weakens by a factor of only
1.5). However, in view of the large and uncertain deadtime effects at
these count rates, we can not exclude that instead the upper peak
becomes much stronger but the entire high-rate spectrum is suppressed
by differential deadtime (van der Klis et al. 1997).  The ratio
between the two kHz QPO powers is less sensitive to dead time (see van
der Klis et al. 1997). For $S_k\geqslant1$ the ratio $P_1/P_2$ of
lower to upper kHz QPO power is lower at high rate sample than at low
rate, but not significantly so (1$\sigma$). At $S_k\geqslant2.5$, the
ratio $P_1/P_2$ is 0.35$\pm$0.08 at low rate, $3.1\sigma$ larger than
the value of 0.02$\pm$0.08 at high rate.

The average count rates and $\nu_2$ values for N=2 using selections
$1.0\leqslant S_k<2.5$ and $S_k\geqslant2.5$ are shown in Fig.\,3.  No
correction for counting-statistics bias was performed; the outer and
inner points should be 6\% and 0.5\% closer together,
respectively. The anticorrelation between $\nu_2$ and the average
count rate is obvious.

It is of interest to compare these short-term correlations to those on
longer time scales. The kHz QPO frequencies vary too little in our
observations ($\nu_2=$1070--1093\,Hz) to determine the flux-frequency
relation on the 10--100~s time scales on which the Z track is traced
out. We can measure this relation in an observation on April 24, 1997,
when Sco X-1 was on its normal branch, covering a wider range of
frequencies ($\nu_2=$1000--1100\,Hz). The NBO was weaker and broader
then. Using 128-sec data segments, we find that also on these longer
time scales, $\nu_{2}$ and count rate are anticorrelated. This is as
expected, as on the normal branch (NB) of Sco X-1 count rate and
position on the Z track are anticorrelated (Priedhorsky et al. 1986;
Dieters \& van der Klis 2000, see also van der Klis 2001). The
fractional frequency shift $\Delta\nu_2/\nu_2$ is 1.51$\pm$0.03 times
the fractional count-rate variation $\Delta C/C$ on these longer time
scales. On the NBO time scale this is 0.43$\pm$0.11, about 1/3 of that
on longer time scales. (Deadtime corrections on $\Delta C/C$ were
neglected here, as both $\Delta C$ and $C$ were nearly equal in the
two cases so that the corrections do not affect the final conclusion.)
In principle this difference could be caused by our method to select
high and low count rates caused by the NBO. However, as we corrected
our results for counting-statistics bias, to the extent that the
dependence of kHz QPO frequency on the NBO flux variations can be
considered linear (which is likely as the variations in NBO amplitude
and kHz QPO frequency are both $<$5\%), these numbers can be directly
compared.  Of course, they were derived from different data sets.

We find that the power ratio $P_1/P_2$, in contrast to what is seen on
the NBO time scale, on long time scales increases with count rate,
from about 0.5 at $\nu_{2}=$~1100 Hz to about 1.0 at $\nu_{2}=$1000 Hz
(this confirms what is indicated by the results shown in figure 3d in
van der Klis et al. 1997; however, in our current analysis, contrary
to that work, we completely cover the 1000--1100\,Hz range without
gaps). The quantity $\Delta(P_1/P_2)/(P_1/P_2)$ is 0.6 times $\Delta
C/C$ on long time scales; on the NBO time scale this is $-$2.

So, a similar anticorrelation between kHz QPO frequency and flux
occurs in Sco X-1 on the NB both on long time scales and on the NBO
time scale, but the weakening of the lower kHz QPO relative to the
upper kHz QPO that we observe towards higher count rate on the NBO
time scale is {\it opposite} to that observed on longer time scales.

\section{Discussion and Conclusion}\label{disc}

In response to the $\sim$6~Hz flux variations due to the NBO in Sco
X-1 the upper kHz QPO frequency $\nu_2$ decreases with count
rate. This is qualitatively, yet not quantitatively consistent with
the behavior of Sco X-1 on the normal branch on longer time scales,
which, however, due to statistics we could only measure using a
different method and in another data set, and at kHz QPO frequencies
including, but mostly below those where we measured the short-term
correlation. The lower kHz QPO weakens relative to the upper one when
flux increases on the NBO time scale. This is {\it opposite} to what
(on the NB for $\nu_2>1000$\,Hz) is observed on longer time
scales. The weakness of the lower kHz QPO in the high count-rate phase
of the NBO cycle limits the accuracy of the determination of its
frequency shifts. Although the frequency separation between the two
peaks appears different between the high count-rate phase and the low
count-rate phase, we can not rule out the possibility that it remains
constant.  In any case, our results show that both frequency and
amplitude of kHz QPOs are affected by flux variations on subsecond
time scales, and that in some respects kHz QPOs react differently to
NBO-related flux variations than to slower ones.

The frequency shift of 22 Hz we observe in the upper kHz QPO peak on
the NBO time scale is of the same order as its width
(FWHM$\sim$70~Hz).  So, the width of the upper kHz QPO peak usually
reported is due in part to short-time-scale variations in kHz QPO
frequency that went unnoticed up to now. As our measurements show
that the kHz QPO signals also change in relative strength on this time
scale, the short-term fluctuations must also affect the $\Delta\nu$
measured on longer time scales: due to what is happening on the NBO
time scale, in power spectra averaged over longer time intervals the
kHz QPO peaks appear closer together than they really are.

The fact that the lower kHz QPO becomes {\it weaker} with respect to
the upper kHz QPO with increasing count rate on the NBO time scale,
but {\it stronger} on longer time scales, strongly suggests that
two different mechanisms produce the count rate variations on these two
time scales, which interact with the kHz QPO production
mechanism in different ways.

The coupling between the kHz QPOs and the NBO provides clues to the
origin of both phenomena. One possibility for the NBO is that it is a
disk oscillation mode in a subsonic region of a thick disk, as
suggested in the sound wave model of Alpar et al. (1992).  In this
model, the NBO frequency is of the same order as the orbital frequency
in a thick disk. If the upper kHz QPO is also related to an orbital
frequency, the NBO is clearly generated at a much larger radius.  The
mass accretion rate in the inner disk region could then be modulated
by this oscillation, and in that way affect $\nu_2$. The same relation
between $\Delta\nu_2/\nu_2$ and $\Delta C/C$ would then be expected to
hold not only qualitatively but also quantitatively on both long time
scales and the NBO time scale, which seems inconsistent with our
results.  However, our measured long-term and short-term values do not
refer to the same data set and were measured using different methods.
Another possibility is that the NBO is associated with a radial
accretion mode, as proposed in the radiation feed-back model (Fortner,
Lamb and Miller 1989, Miller and Park 1995). The X-ray flux modulation
in this model arises either by modulation of the mass inflow rate of
the radial inflow, or by modulation of its Compton scattering
opacity. As shown in Fig.\,3, when $\nu_{2}$ is around 1080 Hz, the
frequency decreases by $\sim$2\% for an increase in average count rate
by $\sim$5\%. From equation 39 in Miller et al. (1998), at an
accretion rate of $\sim0.8$ Eddington and a constant radius near the the
marginally stable orbit (this may approximately represent the
accretion state of Sco X-1 when the NBO is observed), a flux variation
of 6.5\% would yield an orbital frequency change of $\sim$ 5\% due to 
a change in radial radiation force, which
is of the same order of magnitude as observed. Both radiation drag 
and radial force effects would lead to increase radius when 
flux increases. This suggests that the frequency shifts
may be caused by the modulation in the radiative stresses caused by the
NBO. For this the inner boundary of the radial inflow, where it
produces its radiation, should be well inside the inner disk boundary,
some 16\,km for a 1.4 M$_\odot$ neutron star at a $\nu_2$ of
1080\,Hz. This is consistent with the idea that this flow produces its
radiation at or close to the neutron star surface. The quantitative
difference between the $\nu_{2}$ variation on the NBO time scale and
that on longer time scales may come from the fact that the NBO is only
related to the radial inflow, which takes place outside the equatorial
plane, while the long-term variation is related to the accretion flow
through the disk, accreting in the equatorial regions. The radiation
from near the equator may interact more strongly with the inner disk
than that emitted from higher latitudes. A third possibility is that
the NBO originates from the neutron star surface by a mechanism such
as g-modes (Bildsten and Ushomirsky 1996). This can lead to radiative 
stresses which modulate the upper kHz QPO frequency.  Finally,
the Lapidus, Nobili \& Turolla (1994) mechanism for photospheric
oscillations during X-ray bursts, which relies on a near-Eddington
flux, could perhaps also work in Sco X-1 to generate a flux variation
at the NBO frequency emerging from the neutron star surface.

We have found that the flux-frequency correlation on the NBO time
scale and on longer time scales has the same sign in Sco X-1 on the
normal branch: frequency drops when count rate increases. The
situation on Sco X-1's normal branch is actually unusual; what is more
commonly observed (on time scales of hours to days) is that frequency
increases with count rate. It is interesting to ask what would be
observed if in such a case the dependence of kHz QPO frequency on NBO
phase could be measured. If the general idea is correct that the
radiative stresses on the inner disk edge, modulated at the NBO
frequency by some other mechanism than a modulation of the disk
accretion rate, lead to the observed frequency shifts of the upper
kHz QPO, then this anticorrelation would be expected to hold for all
NBO signals produced in the same way as in Sco X-1. The prediction is
that in the more usual situation of a positive long-term correlation
between kHz QPO frequency and count rate, on the NBO time scale there
would still be an anticorrelation, so that then the long and
short-term effects would be of opposite sign. However, there is
evidence that NBOs arise variously as a consequence of luminosity and
radiative-transfer fluctuations (Fortner et al. 1989), so that the
complete picture may be more complex than suggested by this line of
reasoning.

As the frequency correlations between QPOs and noise among neutron
star and black hole X-ray binaries occur over a wide range of source
types and luminosities, further study of how these features are
correlated on short time scales will help reveal the origin of QPOs
and noise components, and understand the tracks in the plot of kHz QPO
frequency vs. X-ray flux. The study of the kHz QPO or high frequency
QPO in response to QPO and noise components at lower frequencies in
black hole and neutron star X-ray binaries are particularly promising
approaches.

\acknowledgments
We are grateful to Jeroen Homan, Mariano M\'endez, Marc Klein-Wolt for
carefully reading the manuscript and useful comments on this work. WY
thanks Eric Ford and Robert Fender for assistances on the use of the
computation facilities in the X-ray group. PGJ was supported by NWO
Spinoza grant 08-0 to E.P.J.van den Heuvel.  WY was supported by NWO
grant 614.051.002 and the National Natural Science Foundation of China
and the Chinese Academy of Sciences. This work has made use of data
obtained through the High Energy Astrophysics Science Archive Research
Center Online Service, provided by the NASA/Goddard Space Flight
Center.

\newpage
\begin{table}
\tabletypesize
\scriptsize
\tablecaption{Log of the Selected Observations}
\begin{tabular}{ccc}
\tableline
Observation ID & Segments & Data Modes\tablenotemark{a} \\
\tableline
10056-01-04-02 & 1,2,3,4 & B+2LLD\tablenotemark{b}\\
20053-01-01-03 & 1,2 & B+SB+2LLD\\
20053-01-01-05 & 3,4 & B+SB+2LLD\\
20053-01-02-03 & 1,2 & B+SB+2LLD\\
20053-01-02-030 & 2,3,4,5 & B+SB+2LLD\\
30035-01-02-00 & 1 & B+SB+2LLD\\
30035-01-02-000 & 4,5 & B+SB+2LLD \\
30035-01-03-00 & 1,2,3,4,5,6 & B+SB+2LLD\\
30035-01-07-00 & 1,2,3 & B+SB+2LLD \\
\tableline
\end{tabular}
\tablenotetext{a}{The time resolution of {\it Bin Mode} (B), {\it
Single-Bit Mode} (SB) and {\it Double Event Mode} (2LLD) data is
2$^{-12}$\,s ($\sim$244
$\mu$s) unless otherwise mentioned. {\it Bin Mode} covers the original PCA
channels 0--49, {\it Single-Bit Mode} 50--249, and {\it Double Event
Mode} 0--249 unless otherwise noted.}
\tablenotetext{b}{The time resolution of {\it Bin Mode} and {\it
Double Event} data is 2$^{-13}$\,s, and the {\it Bin Mode} covers only
PCA channels 0--87.}
\end{table}


\newpage
\begin{figure}
\plotone{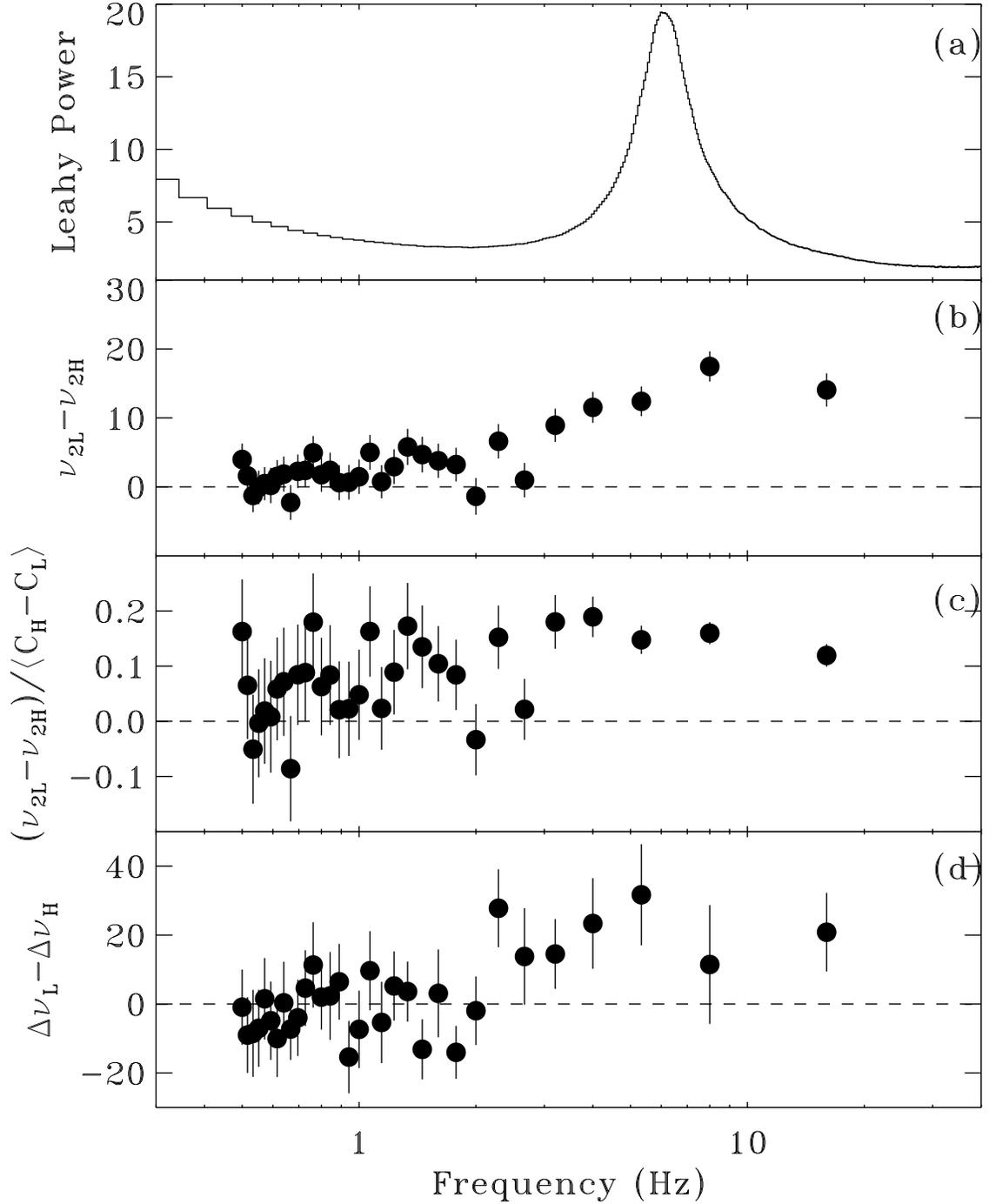}
\caption{Comparison between the high count-rate sample and the low count-rate sample 
with the criterion ${S}_{k}\geqslant1.0$. 
(a) the average power spectrum showing the NBO.  
(b) the frequency difference of $\nu_{2}$. 
(c) the frequency difference over the average count difference in ${2}^{-5}$s
time bin. 
(d) the difference of $\Delta\nu$. }
\end{figure}

\begin{figure}
\plotone{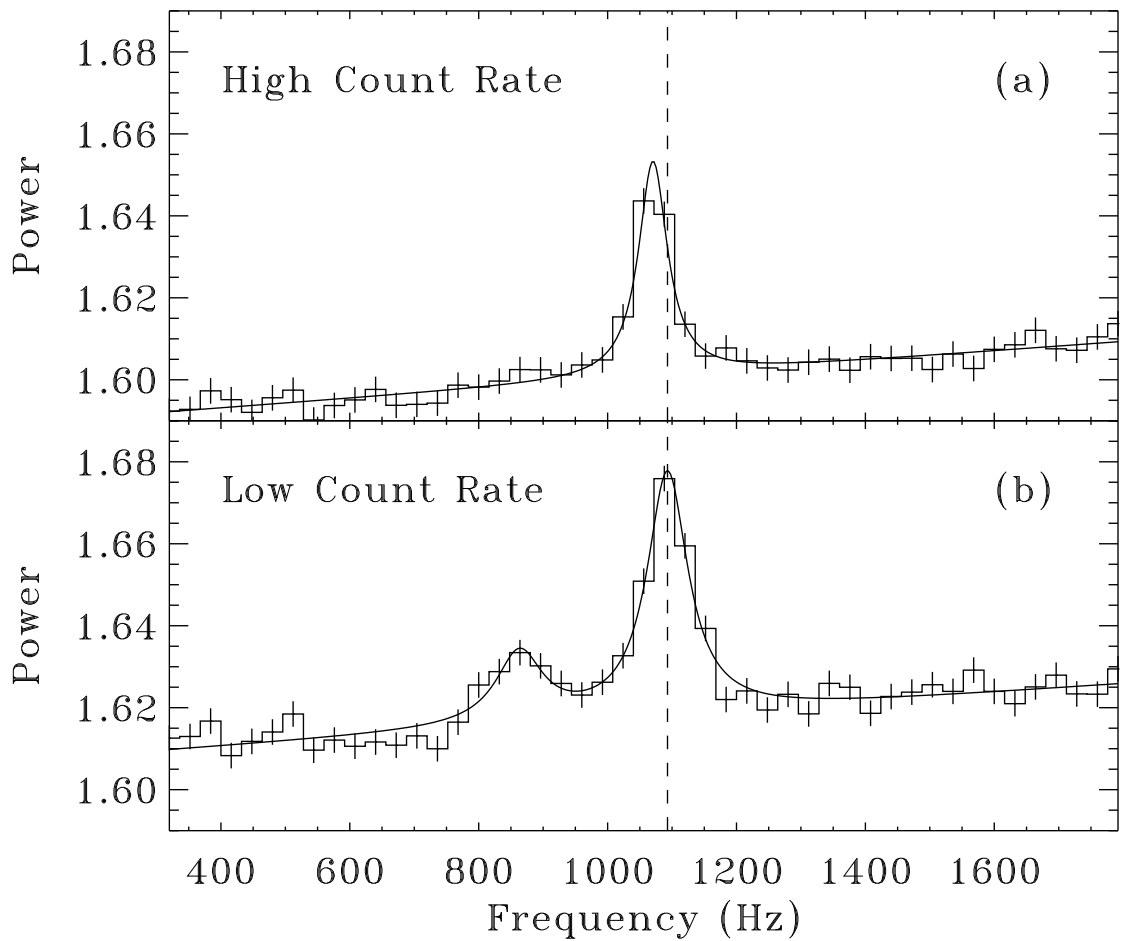}
\caption{The average power spectra of the high count-rate sample and the 
low count-rate sample corresponding to ${S}_{k}\geqslant2.5$ and N=2. 
The dashed line marks $\nu_{2}$ in (b). (a) $\nu_{2}=1070.5\pm1.5$ Hz, FWHM=$54\pm8$ Hz; (b) $\nu_{2}=1092.9\pm2.0$ Hz, 
FWHM=$77\pm6$ Hz.}
\end{figure}

\begin{figure}
\plotone{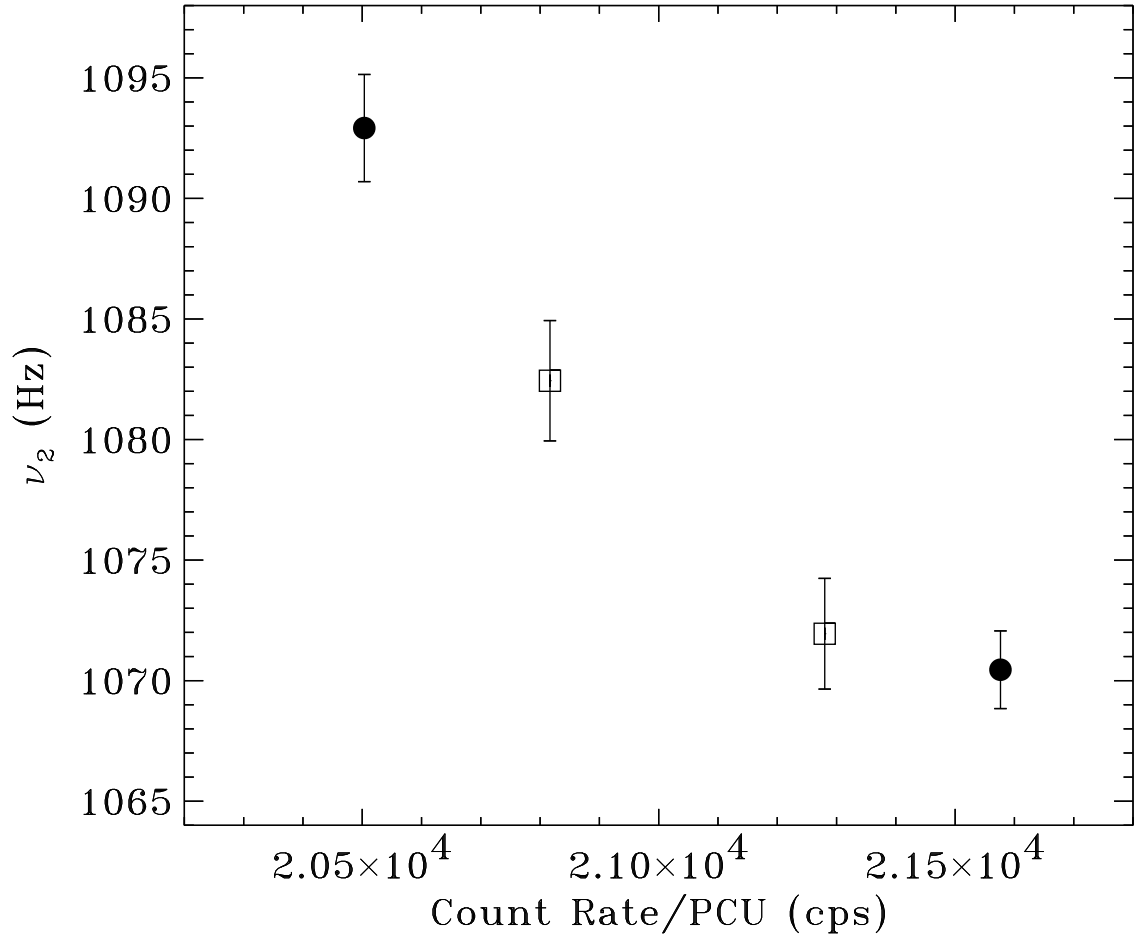}
\caption{The correlation between $\nu_{2}$ and the average count rate
at the NBO frequency (N=2).  Data obtained from
$1.0\leqslant{S}_{k}<2.5$ are marked as squares, and those from
${S}_{k}\geqslant2.5$ are marked as filled circles. Note that contrary
to the rest of this paper, the count rates have not been corrected for
counting statistics bias associated with our selection procedure. }
\end{figure}

\begin{thebibliography}{}
\bibitem[Alpar]{alpar92} Alpar, M. A., Hasinger, G., Shaham, J., \& 
Yancopoulos, S. 1992, A\&A, 257, 627

\bibitem[Bildsten \& Ushomirsky 1996]{bildsten96} Bildsten, L. \& Ushomirsky, G. 1996, \apj,
460, 827

\bibitem[Hasinger et al. 1990]{hasinger90} Hasinger, G., van der Klis, M., Ebisawa, K.,
Dotani, T. \& Matsuda, K. 1990, A\&A, 235, 131
\bibitem[Fortner, Lamb \& Miller 1989]{fortner89} Fortner, B., Lamb, F. K. \& Miller, G.
1989, Nature, 342, 775
\bibitem[Homan et al. 2001]{homan01} Homan, J., van der Klis, M.,
Jonker, P.G., Wijnands, R., Kuulkers, E., M\'endez, M., Lewin, W.H.G.,
2001, ApJ submitted; astro-ph/0104323 
\bibitem[Lapidus, Nobili \& Turolla 1994]{lapidus94} Lapidus, I., 
Nobili, L. \& Turolla, R.  1994, \apjl, 431, L103

\bibitem[M\'{e}ndez 1999]{mendez99} M\'endez, 1999, 19th Texas Symp., astro-ph/9903469

\bibitem[Miller, Lamb \& Psaltis 1998]{miller98} Miller, C. M., Lamb, F. K.
\& Psaltis, D. 1998, ApJ, 508, 791

\bibitem[Miller \& Park 1995]{miller95} Miller, G. S. \& Park, M. 1995, \apj, 440, 771

\bibitem[van der Klis et al. 1996]{klis96} van der Klis, M., Swank, J., Zhang, W.
Jahoda, K. et al. 1996, IAU Circ. 6319

\bibitem[van der Klis et al. 1997]{klis97} van der Klis, M., Wijnands, R. A. D., 
Horne, K. \& Chen, W.  1997, \apjl, 481, L97

\bibitem[van der Klis 2000]{klis00} van der Klis, M. 2000, ARA\&A, to appear in September issue 
(astro-ph/0001167)

\bibitem[van der Klis 2001]{klis01} van der Klis, M. 2001, to appear in ApJ

\bibitem[Wijnands et al. 1997]{wijnands97} Wijnands, R., Homan, J.,
van der Klis, M., M\'endez, M., Kuulkers, E., et al., 1997, \apjl, 490, L157

\bibitem[Wijnands, van der Klis and Rijkhorst 1999]{wijnands99} Wijnands, R.,
van der Klis, Rijkhorst, E.J., 1999, \apjl, 512, L39

\end{thebibliography}
\end{document}